# Rectification of radio frequency current in ferromagnetic nanowire

A. Yamaguchi and H. Miyajima

Department of Physics, Keio University, Hiyoshi 3-14-1, Yokohama, Kanagawa 223-8522, Japan

T. Ono

Institute for Chemical Research, Kyoto University, Gokasho, Uji, Kyoto 611-0011, Japan

Y. Suzuki

Graduate School of Engineering and Science, Osaka University, Machikaneyama 1-3, Toyonaka, Osaka 560-8531, Japan

S. Yuasa, A. Tulapurkar

Nanoelectronics Research Institute, National Institute of Advanced Industrial Science and Technology, Tsukuba 305-8568, Japan

Y. Nakatani

University of Electro-communications, Chofugaoka 1-5-1, Chofu, Tokyo 182-8585, Japan

**[Abstract]**

We report the rectification of a constant wave radio frequency (RF) current by using a single-layer magnetic nanowire; a direct-current voltage is resonantly generated when the RF current flows through the nanowire. The mechanism of the rectification is discussed in terms of the spin torque diode effect reported for magnetic tunnel junction devices and the rectification is shown to be direct attributable to resonant spin wave excitation by the RF current.

Recently, the spin and electric charge of electrons in nano-artificial systems have been seen to exhibit peculiar behavior in the radio-frequency (RF) region through ferromagnetic resonance (FMR), which has led to the spin-polarized current in the RF region becoming attractive from the perspective of a practical utility as a form of RF electric device technology. Investigation of the spin-polarized current has progressed, with applications to spintronics devices, such as magnetoresistive-random-access memories and microwave generators, in mind.

In this study, we present a rectifying effect initially observed in a single ferromagnetic nanowire. The rectifying effect is a consequence of the resonant excitation of spin waves by the RF current. In other words, when the spin wave is excited in the nanowire, the magnetic moment precesses and results in resistance oscillation originating from the anisotropic magnetoresistance (AMR) effect. This oscillation generates the DC voltage by combining with the RF current, as observed in the case of the spin-torque diode effect in the pillar structure systems [1, 2].

We fabricated a system comprising 30 and 50-nm-thick $Ni_{81}Fe_{19}$ nanowires and a 20-nm-thick Au nanowire on MgO substrates, as shown in Fig. 1. The thickness of the Au electrode is 100 nm, while the widths of the 50-nm-thick $Ni_{81}Fe_{19}$ nanowires are 300, 650, and 2200 nm and those of 30-nm-thick $Ni_{81}Fe_{19}$ and Au nanowires are 5 μm and 300 nm, respectively. Considering that the resistivity of $Ni_{81}Fe_{19}$ film is about 26 μΩcm, we designed the lengths of the electrodes gaps so as to match the impedance, with the lengths of the nanowires being far longer than those of the electrodes gaps. A Ground-Signal-Ground (GSG) typed microwave probe is connected by a nanowire, and the bias tee circuit detects the DC voltage difference induced by the RF current flowing through the nanowire. A vector network analyzer injects the constant wave RF current into the nanowire. The external magnetic field $\vec{H}_{ext}$ is applied in the substrate plane as a function of angle $\theta$ to the longitudinal axis of the nanowire, while the Oersted field produced by the RF currents flowing in the loop of the Au electrodes is canceled at the nanowire position because of the symmetrical arrangement of the Au electrode.

The magnetic field dependence of the DC voltage difference generated in the nanowires with widths of 300 nm and 5 μm is shown in Fig. 2 (a) and (b), respectively. The DC resistance of the respective nanowire is (a) 35 Ω and (b) 28 Ω, and the RF current density injected into the nanowire is (a) $6.5 \times 10^{10}$ A/m² and (b) $2.3 \times 10^{10}$ A/m², which represent relatively low values for the Joule heating to be applied. An external magnetic field is applied in the substrate plane at angles of $45°$ and $45° + 180° = 225°$ to the wire-axis. As seen in the figure, the peak position of resonance spectrum shifts to a higher frequency region with the external magnetic field. Surprising is the fact that the sign of the DC voltage reverses when the sense of the applied magnetic field is reversed, which corresponds to the case wherein the angle is changed from $\theta$ to $\theta + 180°$. On the other hand, no similar effects are observed in the 20-nm-thick Au nanowires, which strongly indicates that the resonance is of magnetic origin and generated by the RF current.

The magnetic field dependence of the resonance frequency of the nanowire with widths of 300, 650, 2200, and 5000 nm, where the magnetic field is applied at an angle of $45°$ to the wire-axis. The resonance frequency increases with increasing magnetic field but decreasing wire-width, which means that the resonance mode possibly reflects the shape magnetic anisotropy.

According to the extended Kittel's equation [3, 4], the ferromagnetic resonance frequency $f = 2\pi/\omega$ at the magnetic field $H$ is approximately given by

$$f(H) = \frac{g\mu_B\mu_0}{h} \cdot \left[\left(H + H_A + M_S\left(\frac{1-e^{-kd}}{kd}\right)\right) \cdot (H + H_A)\right]^{\frac{1}{2}}, \tag{1}$$

where $M_S$ denotes the saturation magnetization, $g$ the Lande factor, $k$ the wave number of the spin wave, $d$ the thickness of the sample, and $H_A$ the effective anisotropy field including demagnetizing and anisotropy fields in the substrate plane. The experimentally determined $H_A$ for the nanowires with widths of 300, 650, 2200, and 5000 nm is 1160, 830, 260, and 100 Oe, respectively. The experimentally obtained resonance frequency correlates well to calculated by the extended Kittel's eq. (1), indicating the existence of resonant spin wave excitation. The precession of the magnetic moments in the nanowires, meanwhile, which occurs due to the spin wave excitation caused by the RF current, results in resistance oscillation at the resonance frequency due to the AMR effect. The additional wiggles in the trace of Fig. 2 corresponds to the additional spin wave modes. This resistance oscillation generates the DC voltage by combining with the RF current, as observed in the case of the spin-torque diode effect in the pillar structure systems [1, 2]. Next, we derive an analytical expression for the DC voltage produced in the nanowire.

The RF-current-induced dynamics of the magnetic moment are analytically calculated by the modified Landau-Lifshitz-Gilbert (LLG) equation, including the spin-transfer term [5 - 7], The LLG equation in the coordinate system shown in Fig. 4 (a) is expressed by

$$\frac{\partial \vec{m}}{\partial t} = -\gamma_0 \vec{m} \times \vec{H}_{eff} + \alpha \vec{m} \times \frac{\partial \vec{m}}{\partial t} - (\vec{u}_s \cdot \nabla) \vec{m}, \tag{2}$$

where $\vec{m}$ denotes a unit vector along the local magnetization, $\gamma_0$ the gyromagnetic ratio, $\vec{H}_{eff}$ the effective magnetic field, including the exchange and demagnetizing fields, and $\alpha$ the Gilbert damping constant respectively. The last term in eq. (2) represents the spin-transfer torque, which describes the spin transferred from conduction electrons to localized spins. The spin-transfer effect is a combined effect of the spatial nonuniformity of magnetization and the flowing current. A vector with a dimension of velocity, $\vec{u}_s = -\vec{j} P g \mu_B / (2 e M_s)$, is essentially the spin current associated with the electric current in a ferromagnet, where $\vec{j}$ is the current density, $P$ the spin polarization of the current, and $e$ the electronic charge respectively. The spatial nonuniformity of the magnetization in a nanowire is mainly caused by the spatial dispersion of the demagnetizing field.

Now, suppose we have an (x, y, z) coordinate system: each component corresponds to the short wire axis, longitudinal wire axis, and normal to the substrate plane, respectively, as schematically shown in Fig. 3 (a). The external magnetic field is applied along an angle $\theta$ from the y-coordinate axis. Then, let us define a coordinate system (a, b, c) where the b direction corresponds to the equilibrium direction of $\vec{m}_0$ along the effective magnetic field $\vec{H}_{eff}$, assuming for the sake of simplicity that $\vec{H}_{eff}$ is almost parallel to the external magnetic field. This assumption is valid provided the spin wave excitation is small and the external magnetic field is lower than the effective anisotropy field. A small variation $\delta \vec{m} = (m_a, 0, m_c)$ in the effective magnetization $\vec{m}_0 = (0,1,0)$ is obtained by the Laplace transformation, considering that $\vec{m} = \vec{m}_0 + \delta \vec{m}$ in the (a, b, c) coordinate system. The magnetization variation $\delta \vec{m} = (m_x, m_y, m_z)$ in the (x, y, z) coordinate system can be obtained by a rotation transformation from the (a, b, c) coordinate system to the (x, y, z) coordinate system;

$$\begin{pmatrix} m_x \\ m_z \end{pmatrix} = \frac{j}{(\omega_k + i\omega\alpha)^2 - \omega^2} \begin{pmatrix} i\omega & \omega_k + i\omega\alpha \\ -(\omega_k + i\omega\alpha) & i\omega \end{pmatrix} \begin{pmatrix} s_a \cos\theta \\ s_c \end{pmatrix}, \qquad (3)$$

where $\omega_k = 2\pi f(H)$ denotes the spin wave resonant angular frequency corresponding to $f(H)$ given by eq. (1), and $S_a$ and $S_c$ denote the components of the spin transfer torque $\vec{S}$ defined as $\vec{s} = (\vec{u}_s \cdot \nabla)\vec{m} = j \cdot (s_a, s_b, s_c)$ in the (a, b, c) coordinate system.

The small variation in the magnetic moment around the direction of the effective magnetic field, due to the spin wave excitation at a precession angle of $\phi_l$, generates the time-dependent AMR effect. When $\cos^2 \phi_l \approx 1$ and $\sin^2 \phi_l \approx 0$ for small $\phi_l$, we can approximately derive the time-dependent AMR $\langle R(t) \rangle = \left\langle \sum_l \Delta R \cos(\theta + \phi_l(t)) \right\rangle$, which means the average of the AMR $R(t)$ in each cell, by using the additional formula for cosine:

$$\langle R(t) \rangle \approx \Delta R \left[ \cos^2 \theta - \left\langle \frac{1}{2} \sin 2\theta \sin 2\phi_l(t) \right\rangle \right], \tag{4}$$

where $\Delta R$ denotes the resistance change due to the AMR effect when the angle between the current and the magnetization changes from $0°$ to $90°$, $l$ the index of the numbered domain of the spin wave excitation in the wire, and $\sin 2\phi_l(t) \approx 2[\Delta \vec{m}/|\vec{m}|]_x$, the x component of $\delta \vec{m}$. When the RF current with amplitude $I$ is written by $I(t) = I \cos(\omega t)$, the frequency variation of the induced voltage is given by the Laplace transformation of $V(t) = R(t) \cdot I(t)$:

$$V(\omega) = A(\omega) \cdot I^2 \cdot \sin 2\theta \cos \theta, \tag{5}$$

where $A(\omega)$ is expressed by

$$A(\omega) \approx \frac{2\omega^2 \alpha \omega_k \cdot s_a + (\omega_k^2 - \omega^2) \omega_k \cdot s_c}{(\omega_k^2 - \omega^2)^2 + 4\omega^2 \alpha^2 \omega_k^2} \cdot \frac{\Delta R}{b \cdot d}, \tag{6}$$

where $b$ and $d$ are the width and thickness of the nanowire, respectively. Figure 3 (b)

shows the amplitude of the DC voltage at the resonance frequency as a function of $\theta$. The solid line shows the angular dependence obtained by eq. (5) and is in good agreement with the experimental results, while the inset shows the variation of the amplitude of the DC voltage against the square of the RF current in $H$= 400 Oe and $\theta = 35°$. As predicted by eq. (5), the voltage amplitude is proportional to the square of the RF current. However, in the external magnetic field, which is lower than the effective anisotropy field, the experimental results disagree with the calculation predicted by eq. (5) [8].

A micromagnetic simulation of the system, which takes into account the spin transfer torque, is performed to confirm the generation of spin wave excitation by the constant wave RF current [8]. The stripe magnetization pattern induced by the RF current is clearly confirmed, indicating the spin wave excitation. It should be noted that the period of the stripe pattern of approximately 100 nm is consistent with the spin wave length estimated by $\sqrt{A/K}$, where $A$ and $K$ are the exchange stiffness constant and the anisotropy energy written by $K = g\mu_B\mu_0 H_A$, respectively.

The rectification of the RF current results in a highly-sensitive measurement of spin dynamics in nano-scale magnets. As the effect can be easily tuned to control the resonance frequency, based on the shape of the wire or the magnetic field, it will facilitate the development of spintronics devices.

We would like to thank H. Kohno at Osaka University for useful discussions. The present work was partly supported by the Keio Leading-edge Laboratory of Science and Technology project 2006.

**Figure caption**

Figure 1

The RF electric circuit and the optical micrograph (top view) of the system consisting of electrodes and the magnetic nanowire.

Figure 2

DC voltage generated by (a) the 50-nm-thick $Ni_{81}Fe_{19}$ wire with width of 300 nm and (b) the 30-nm-thick $Ni_{81}Fe_{19}$ wire with width of 5 μm in response to the RF current. Each resonant response is vertically shifted for clarity.

Figure 3

(a) Schematic coordinate system adopted in this calculation. The longitudinal wire axis is parallel to the y-axis. (b) Angle dependence $\theta$ of the DC voltage amplitude. The dashed-line represents a $\sin 2\theta \cos\theta$ curve. The inset shows the linear relationship between the amplitude and the square of the RF current.

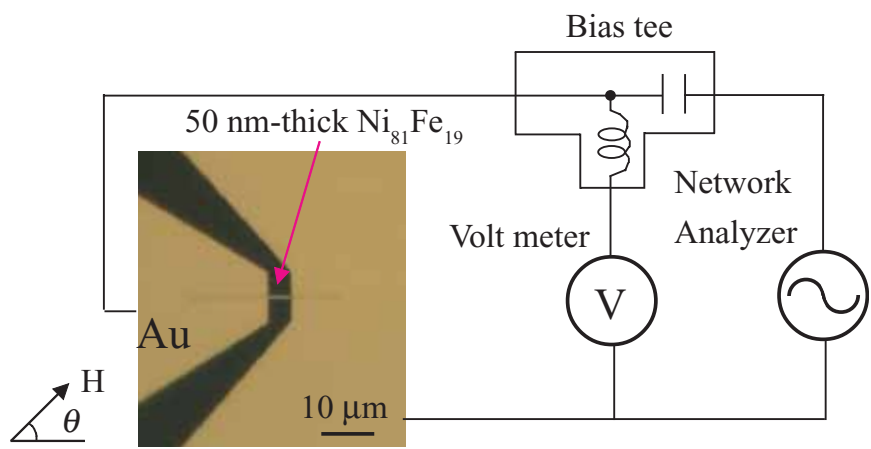

Fig. 1 A. Yamaguchi *et al*.

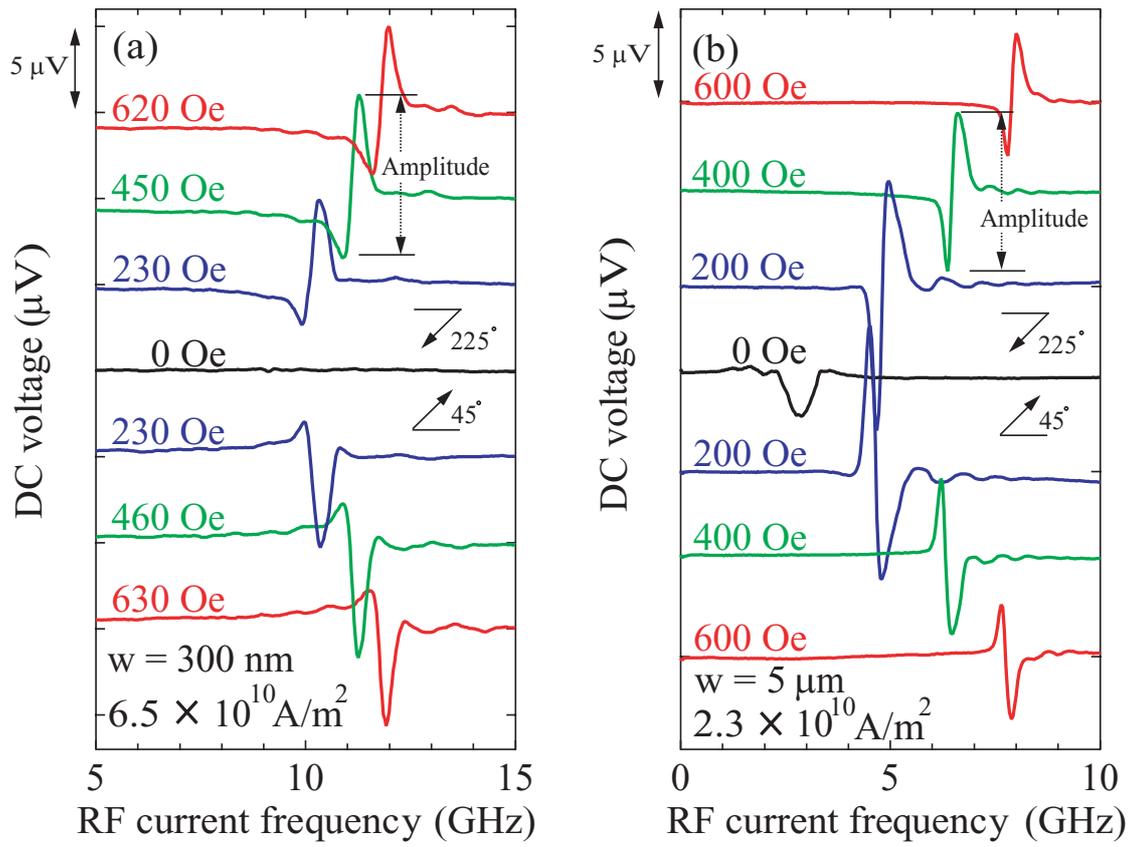

Fig. 2 A. Yamaguchi *et al*.

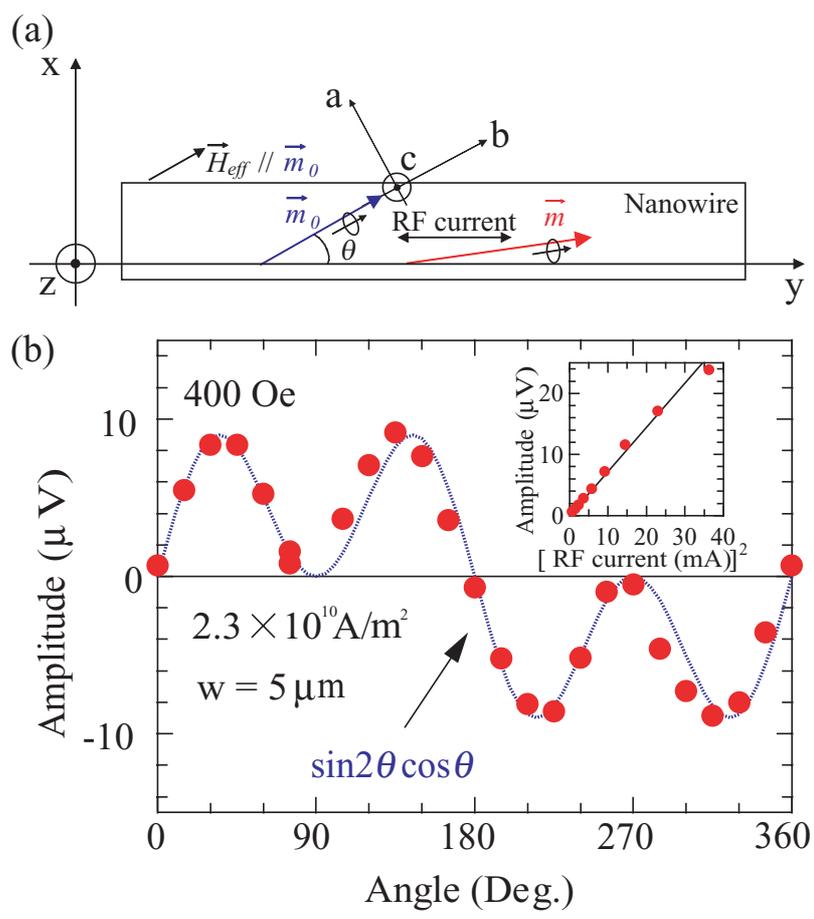

Fig. 3 A. Yamaguchi *et al*.